\documentclass[manuscript]{aastex}

\usepackage{epsfig}
\input psfig.sty

\usepackage{graphicx}
\usepackage{dcolumn}
\usepackage{bm}
\usepackage{epsfig}
\usepackage{graphicx}
\usepackage{dcolumn}
\usepackage{bm}

\shorttitle{Influence of Small-Scale Inhomogeneities on the Cosmological Tests}
\shortauthors{Busti \& Lima}

\begin{document}

\title{Influence of Small-Scale Inhomogeneities on the Cosmological Consistency Tests}

\author{V. C. Busti}
\affil{Instituto de Astronomia, Geof\'isica e Ci\^encias Atmosf\'ericas (IAG-USP), Universidade 
de S\~ao Paulo, 05508-900 S\~ao Paulo, SP, Brazil}
\email{vcbusti@astro.iag.usp.br}

\and

\author{J. A. S. Lima}
\affil{Instituto de Astronomia, Geof\'isica e Ci\^encias Atmosf\'ericas (IAG-USP), Universidade 
de S\~ao Paulo, 05508-900 S\~ao Paulo, SP, Brazil}
\email{limajas@astro.iag.usp.br}

\begin{abstract}
The current cosmological dark sector (dark matter plus dark energy) 
is challenging our comprehension about the physical processes taking 
place in the Universe. Recently, some authors tried to falsify the 
basic underlying assumptions of such dark matter-dark energy 
paradigm.  In this Letter, we show that oversimplifications of the 
measurement process may produce false positives to any consistency 
test based on the globally homogeneous and isotropic $\Lambda$CDM 
model and its expansion history  based on distance measurements.
In particular,  when local inhomogeneity effects due to 
clumped matter or voids are taken into account, an apparent 
violation of the basic assumptions (``Copernican Principle'') seems 
to be present. Conversely, the amplitude of the deviations 
also  probes the degree of reliability underlying the  phenomenological Dyer-Roeder 
procedure by confronting its predictions with the accuracy of the weak lensing approach. 
Finally, a new method is devised to reconstruct 
the effects of the inhomogeneities in a $\Lambda$CDM model, and some 
suggestions of how to distinguish between clumpiness (or void) 
effects from different cosmologies are discussed. 
\end{abstract}


\keywords{cosmological parameters -- cosmology: observations -- cosmology: theory -- dark energy -- large-scale structure of Universe -- gravitational lensing}

\section{Introduction}

The dark energy mystery has inspired cosmologists to test all the 
assumptions of the so-called cosmic concordance model 
($\Lambda$CDM). In the last few years, some methods to detect 
possible deviations from the FLRW (Friedman -- Lema\^itre -- Robertson 
-- Walker) metric \citep{clark08,ellis2008} or the flat $\Lambda$CDM 
model \citep{sss2008,zc2008}, as well as to reconstruct the dark 
energy equation of state $w(z)$ have been proposed  \citep[e.g.][]{ex1,ex2,ex3}.   
However, for a real understanding of what is being measured, it is 
fundamental to check whether such proposals are based on 
assumption-free approaches or whether such deviations are naturally 
mimicked when a more realistic description is considered. 

On the other hand,  the observed universe must be studied in two 
separated spatial regimes since it is homogeneous on large scales ($ 
> 100$ Mpc) while a hierarchy of structures involving 
galaxies, filaments, clusters of galaxies and voids is seen on small 
scales. Such inhomogeneities can change the observed distances when 
radiation is used, because the light rays probe the local 
gravitational field thereby affecting the cosmological parameters. 

In principle, even assuming that the FLRW metric is adequate to 
describe the cosmic expansion history, the existing observations may 
prefer underdense lines of sight as compared to the background, and, 
as such, the distance relations need to be corrected for the 
realistic clumpy Universe. The basic consequence is that artifacts 
(false positives) will be produced in the existing tests originally 
proposed within the globally smooth FLRW model. 
Reciprocally, since the magnitude of the artifacts is heavily dependent on how light propagation is described in the clumpy 
Universe, such tests can also unveil the most suitable method to deal with the inhomogeneities.

The purpose of this Letter is threefold: first, we show that 
small-scale inhomogeneities affect the distance and  produce false 
positives for two distinct tests, namely: the ${\cal{C}}(z)$ 
\citep{clark08} and ${\cal{L}}(z)$ tests \citep{sss2008,zc2008}. 
Second, a new method is proposed to reconstruct the effects of the 
inhomogeneities directly from observational data when a $\Lambda$CDM 
model is assumed, and, finally,  a discussion is performed of how to 
distinguish between the clumpiness (or void) effects  from different 
cosmologies.

\section{Cosmological Tests}

In what follows, we restrict our attention for  the two above quoted cosmological tests (${\cal{C}}(z)$ and ${\cal{L}}(z)$). 
However, it is important to 
stress that any test based on distance measurements will produce the artifacts discussed in this Letter. For
instance, the influence of the inhomogeneities on the reconstruction of the dark energy equation of state 
was discussed by \cite{bolejko2011w}, and a Copernican test involving the 
redshift drift and distances will be affected
as well \citep{ellis2008}.

\subsection[]{The ${\cal{C}}(z)$ Test}

The leitmotiv of such an approach is to test the so-called 
``Copernican Principle'' (CP) which is implicit in the homogeneous 
and isotropic FLRW metric \citep{clark08}. In this case, the 
possible redshift dependence of the curvature parameter (a CP 
violation signature) can be discussed based on the  expression of 
the luminosity distance (in our units $c=1$): 

\begin{equation}
d_L(z)=\frac{(1+z)}{H_0 \sqrt{-\Omega_k}} 
\sin{\left(\sqrt{-\Omega_k}\int_{0}^{z}dz^{\prime}\frac{H_0}{H(z^{\prime})}\right)}, 
\end{equation}  
where $H(z)$ is the expansion rate ($H_0$ is the   
Hubble constant)  and $\Omega_k$ is the present day curvature parameter. By defining $D(z)=\frac{H_0 d_L(z)}{(1+z)}$, one can 
differentiate the above equation and rearrange the terms in order to 
have 

\begin{equation}\label{Omega} 
\Omega_k=\frac{[E(z)D^{\prime}(z)]^2 -1}{D(z)^2},  
\end{equation}
where the prime denotes redshift differentiation and $E(z)=H(z)/H_0$. Now, it is easy to see that a differentiation of (\ref{Omega}) yields 

\begin{equation}
 {\cal{C}}(z)=1+E^2(DD^{\prime\prime}-D^{\prime 2}) + 
EE^{\prime}DD^{\prime} \equiv 0, \label{Cz} 
\end{equation}
since it has been assumed that $\Omega_k$ is constant for all redshifts when the globally smooth FLRW metric properly describes the 
background geometry. It is worth notice that deviations of the order of $10^{-5}$ are expected in 
realistic models due to perturbations related to structure formation for all redshifts \citep{rev_ellis}. 

\subsection{The ${\cal{L}}(z)$ Test}

 Unlike the ${\cal{C}}(z)$ test, the basic  aim here is to 
identify any deviation from  a flat $\Lambda$CDM model. It was 
independently introduced  by \cite{sss2008} and \cite{zc2008}, and may be interpreted as a kind of 
consistency check. For a flat $\Lambda$CDM model, the present value 
of the matter density 
parameter, $\Omega_m$, can be written in terms of the observed quantities: 

\begin{equation}
\Omega_m = \frac{\left[H(z)/H_0\right]^2 -1}{(1+z)^3 
-1}=\frac{1-D^{\prime}(z)^2}{[(1+z)^3 -1]D^{\prime}(z)^2}. 
\end{equation}

Now, following the same approach of the ${\cal{C}}(z)$ test, a simple 
differentiation provides 

\begin{equation}
{\cal{L}}(z)=2[(1+z)^3 -1]D^{\prime\prime}(z) 
+3(1+z)^2D^{\prime}(z)[1-D^{\prime}(z)^2], \label{Lz}  
\end{equation}
which must 
also be identically null (regardless of the redshift) for all flat 
$\Lambda$CDM models. Since the quantities appearing in (\ref{Cz}) 
and (\ref{Lz}) are measurable, one may expect that both null results 
can be checked by the available data. 

\section{The Dyer-Roeder Approach}

In the above discussed tests the Universe was assumed to be 
homogeneous and isotropic on all scales. Therefore, the basic 
question now is: How the background cosmological tests are affected 
by the small-scale structures? In other words,  even assuming that 
the large-scale homogeneity is preserved, the light propagation is 
perturbed by the small-scale inhomogeneities, potentially modifying 
the angular diameter and luminosity distances. Therefore, it is 
fundamental to quantify the unknown physical conditions along the 
light path. 

Initially, this issue was addressed by \cite{Ze64}, 
followed by \cite{bert66}, \cite{gunn67} and \cite{Kant69}. But in the beginning of 70's  
\cite{Dy72,Dy73} adopted the average path assumption so that the 
underdensities in voids are compensated by overdensities in clumps, 
thereby making the universe homogeneous only on very large scales. A 
typical line of sight is far from the clumps, not suffering from 
gravitational lensing effects. In this way, the unknown  physical 
conditions along the path, associated with  the clumpiness effects, 
are phenomenologically described by the smoothness 
parameter $\alpha$. Such a quantity has a straightforward physical meaning: 
it is the fraction of homogeneously distributed matter within a 
given light cone. For $\alpha=0$ (empty beam), all matter is clumped 
while for $\alpha=1$ the fully homogeneous case is recovered. Then, 
for a partial clumpiness, the standard interpretation (involving 
structures more massive than the cosmic average) is that the 
smoothness parameter is restricted only over the interval $[0,1]$. 

Observationally, the smoothness parameter is still poorly 
constrained. By using compact radio sources, no constraint over 
$\alpha$ was obtained \citep{AL04,SL07}, whereas an analysis with 
Type Ia Supernovae (SNe Ia) in a flat $\Lambda$CDM model constrained 
$\alpha \ge 0.42$ within the 95.4\% confidence level ($2\sigma$) 
\citep{SCL08}. The introduction of $H(z)$ data only mildly improved 
the results: $\alpha \ge 0.66$ within the 95.4\% confidence level 
\citep{hzdata}. In the same vein, by combining the 557 SNe Ia from the Union 2 compilation \citep{union2} 
and 59 Gamma-Ray Bursts compiled by \cite{hymnium}, it was shown 
that $\alpha \ge 0.52$, i.e. a more inhomogeneous Universe
is compatible with current data (Busti, Santos \& Lima 2012).

Several generalizations of the Dyer-Roeder approach have been 
proposed in the literature. The dependence of the smoothness 
parameter with the redshift was first discussed  by  
\cite{linder88,linder98} and  \cite{SL07}. The influence 
of a non-standard expansion rate was analysed by  
\cite{mattsson2010}, and a connection with weak lensing was also 
investigated by  \cite{bolejkodr2011}. A comprehensive study 
concerning inhomogeneity 
effects on light propagation was recently carried out by \cite{clark11}. 
The Dyer-Roeder approach we chose to deal with
the inhomogeneities is not unique. There are other proposals in the 
literature, e.g. \cite{marra2009}, but the simplest one is the 
Dyer-Roeder approach.  

The above discussions reveal that the small-scale inhomogeneities 
affect the light propagation although its modeling is far from trivial. 
Potentially, the inhomogeneities may play an important role, thereby 
masking  several proposed consistency checks of $\Lambda$CDM and 
other dark energy models. Therefore, in order to claim a violation 
of the CP (${\cal{C}}(z)$ test) or any deviation from a flat $\Lambda$CDM model 
(${\cal{L}}(z)$ test)  it is vital to disentangle all the potential 
effects.

\subsection{The Dyer-Roeder Distance}

The derivation follows from  Sach's optical equation 
\citep{Sachs61,Sachs2} 

\begin{eqnarray}\label{sachs} 
{\sqrt{A}}'' +\frac{1}{2}R_{\mu \nu}k^{\mu}k^{\nu} \sqrt{A}=0, 
\end{eqnarray} 
where a prime denotes differentiation w.r.t. the affine parameter 
$\lambda$, $A$ is the cross-sectional area of the light beam, 
$R_{\mu\nu}$ the Ricci tensor, $k^{\mu}$ the photon four-momentum, 
and the shear was neglected. 

Five steps are needed to achieve the luminosity distance in the 
Dyer-Roeder approach: (i) the assumption that the angular diameter 
distance $D_A \propto \sqrt{A}$, (ii) the relation between the Ricci 
tensor and the energy-momentum tensor through the Eintein's field 
equations, (iii) the relation between the affine parameter $\lambda$ 
and the redshift $z$, (iv) the {\it ansatz} $\rho_m$ goes to $\alpha 
\rho_m$, and (v) the validity of the duality relation between the 
angular diameter and luminosity distances 
\citep{ETHER33,Basset04,Holanda10,Holanda11}. 

For a XCDM model, one obtains the Dyer-Roeder distance $(d_L=H_0^{-1} D_L)$ by solving the equation: 

\begin{eqnarray} 
\frac{3}{2}  \left[ \alpha(z) \Omega_m (1+z)^3 + \Omega_X (1+w)(1+z)^{3(1+w)}  \right]  D_L(z) +  \nonumber 
\\ (1+z)^2  E(z) \frac{d}{dz} \left[ (1+z)^2 E(z) \frac{d}{dz} \frac{D_L(z)}{(1+z)^2}  \right]   =  0,   
\label{angdiamalpha} 
\end{eqnarray} 
where $\Omega_X$, $w$, are the density and equation of state parameters of dark energy while  the dimensionless Hubble parameter, $E(z)= H/H_0$, reads: 
\begin{equation} 
E(z)= \sqrt{\Omega_m (1+z)^3 + \Omega_X (1+z)^{3(1+w)} + \Omega_k(1+z)^2}, 
\end{equation} 
where $\Omega_k=(1-\Omega_m - \Omega_X)$. The initial conditions to solve Eq. (\ref{angdiamalpha}) are: $D_L(0)=0$ and $\frac{dD_L}{dz}|_{0}=1$.

\section{Results}

\subsection{Quantifying the Influence of $\alpha$}

In order to quantify the effects of the inhomogeneities on the 
cosmological tests (${\cal{C}}(z)$  and ${\cal{L}}(z)$) we plot 
expressions (\ref{Cz}) and (\ref{Lz}) by using the Dyer-Roeder distance with 
the following prescription for the smoothness parameter: 

\begin{equation}
\alpha(z) = 1 + {\beta a^{3\gamma}}=1 + 
{\beta(1+z)^{-3\gamma}}, \label{alphaz}  
\end{equation}
where $\beta$ and 
$\gamma$ are constant parameters, and $a\equiv(1+z)^{-1}$ is the 
cosmic scale factor. Since the degree of homogeneity is higher in 
the distant past, it follows that $\gamma \geq 0$ because the limit 
$\alpha$ $\rightarrow 1$ at high redshifts must be obeyed. For a 
given value of $\gamma$, the $\beta$ parameter quantifies  the 
influence of the structure formation process at lower redshifts. For 
the sake of generality, we also consider that $\beta$ (to be fixed 
by the data) may assume  negative and positive values in order to 
describe clumps and voids, respectively. The above deformation of 
the standard  FLRW description ($\alpha=1$) parametrize our 
ignorance on the late time structure process. It is clearly inspired 
by similar expressions for the $\omega(z)$-equation of state 
parameter of dark energy models \citep{Pad03,Lin03}. 
Interestingly, such prescription for $\alpha$ produces the 
same results as weak lensing when the parameters are tuned to
$\beta \sim O(10^{-3})$ and $\gamma=5/12$ \citep{bolejkodr2011}.

\begin{figure*} 
\centerline{\epsfig{figure=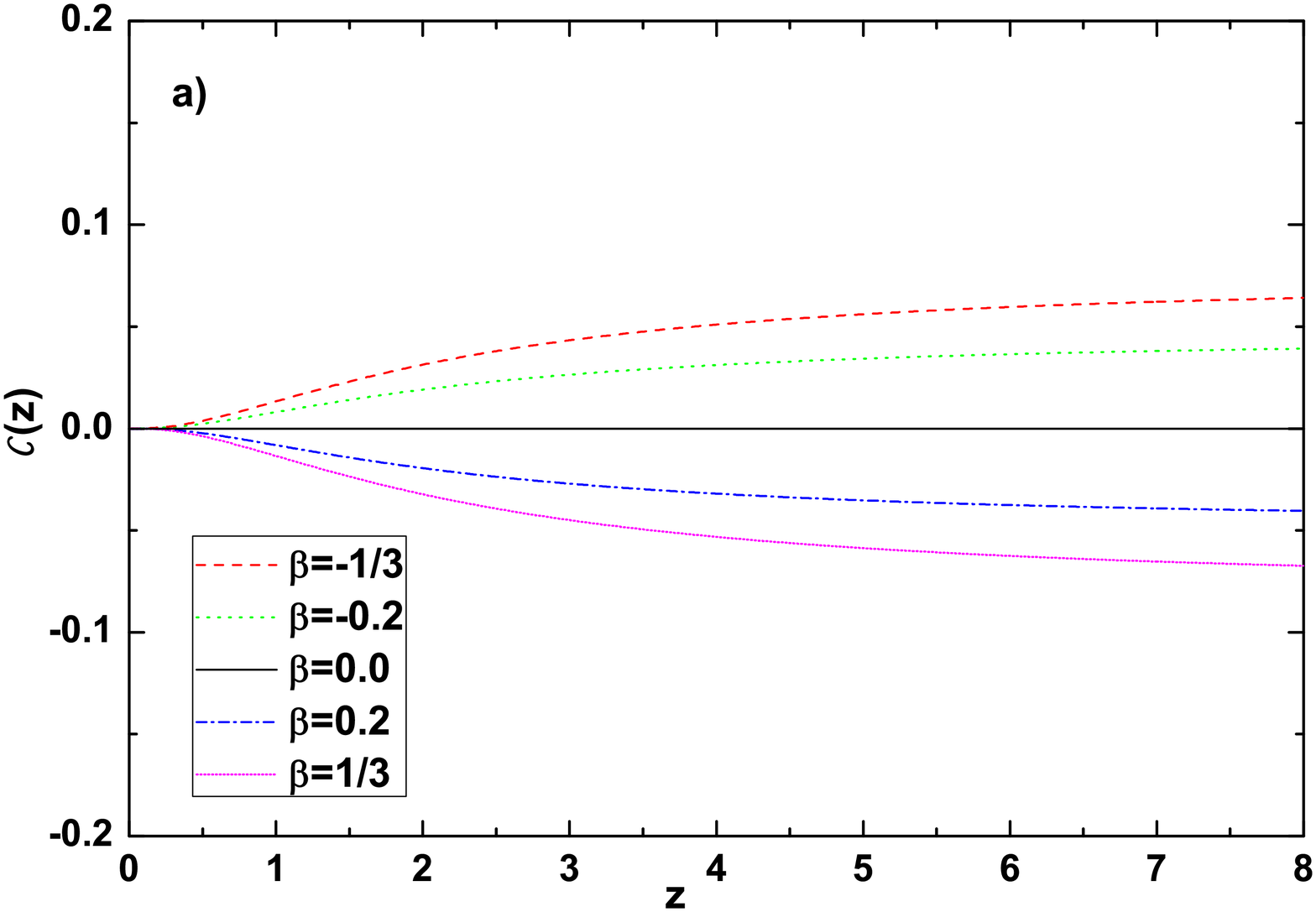,width=3.1truein,height=2.5truein} 
\epsfig{figure=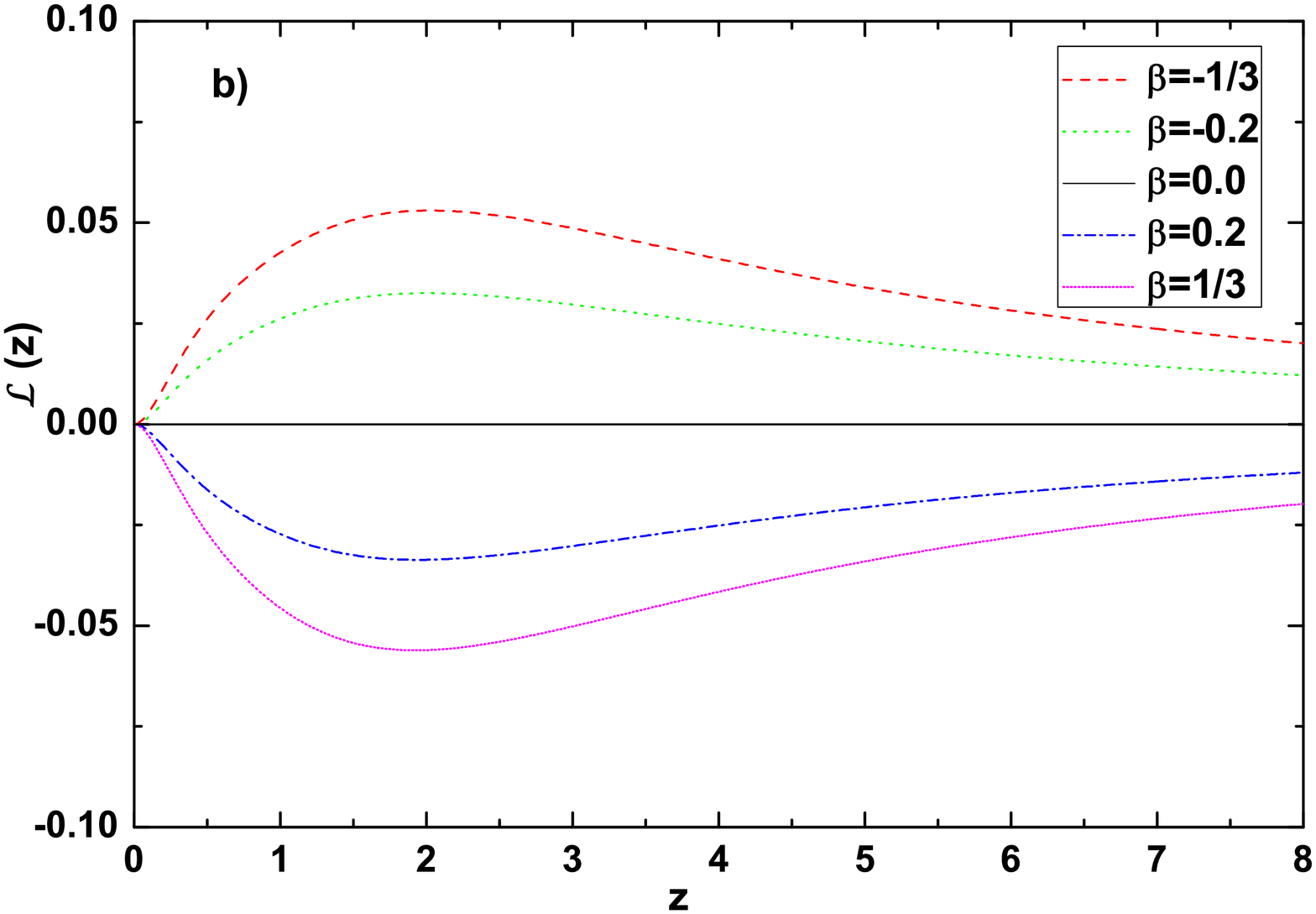,width=3.1truein,height=2.5truein} \hskip 0.1in} \caption{The $\beta$-effect on the 
${\cal{C}}(z)$ and ${\cal{L}}(z)$ cosmological tests. {\bf a)} 
${\cal{C}}(z) \equiv 0$ (solid black line, $\beta = 0$, $\alpha = 1$) is 
the prediction of the standard globally smooth $\Lambda$CDM model [see Eq. 
(\ref{alphaz})]. All curves with $\beta \neq 0$ are simulating an 
apparent violation (false positive) of the Copernican Principle. 
{\bf b)} As in Fig. {\bf a}, ${\cal{L}}(z) = 0$ is the $\Lambda$CDM 
prediction with no inhomogeneous corrections. As  physically 
expected, due to the choice of the $\beta$ values, the corrections 
for clumps ($\beta < 0$, $\alpha < 1$) and voids ($\beta 
> 0$, $\alpha 
> 1$) are symmetric with respect to the $\Lambda$CDM prediction.} \label{figCL} 
\end{figure*}

In Fig. \ref{figCL}, we display the results for $\gamma=1$ and 
four values of $\beta$, where we fixed a flat $\Lambda$CDM model 
with $\Omega_m=0.27$. In the left panel the ${\cal{C}}(z)$ test in 
function of the redshift shows that when local inhomogeneities are 
taking into account deviations from zero are expected even with no 
violation of the FLRW metric. For comparison we also plotted the 
standard case ($\alpha=1$). The magnitude of the effect is 
dependent of the chosen parameters. In this concern, we recall that deviations of the same order of magnitude as displayed in Fig. \ref{figCL}  
were also obtained from void  models constrained by  SNe Ia and $H(z)$ data \citep{render}. In particular, when the free parameters are tuned by the weak lensing prediction, 
the deviation is smaller  ($\sim 10^{-4}$) but higher than expected
from structure formation process alone, which is $10^{-5}$ \citep{rev_ellis}. 

On the other hand,  since the  magnitude of the deviation  depends 
on how light propagation occurs, our approach can be used in the inverted manner. More precisely, it is also suitable to probe 
not only the consistency of the phenomenological Dyer-Roeder approach but also the accuracy of the weak lensing formalism. In the latter case,  caution with the weak lensing approximation should be signalized by higher values 
in the ${\cal{C}}(z)$ test.

In the right panel of Fig. \ref{figCL}, the same analysis is 
performed for the ${\cal{L}}(z)$ test. It is evident that the 
deviations are bigger when compared to the ${\cal{C}}(z)$ test in 
the redshift region $[0,1]$, so it is expected that the effect will 
be measurable by SNe Ia.  By assuming the expected values from weak lensing, the deviations are again around $10^{-4}$.  
One question arises whether the inhomogeneities are playing 
the role of curvature or a non-$\Lambda$ behavior. It is important to remark that 
only the distances are affected by the inhomogeneities, not the expansion rate. 
So a cross-check with ${\cal{L}}(z)$ involving only $H(z)$ and its derivatives is demanded in order to ascribe
what is the primary cause.
In principle, the amplitude of the deviations 
can be used to decide which is the more realistic description of the inhomogeneities. Deviations around $10^{-4}$ is an indication 
favouring  weak lensing. However, whether higher values are obtained the phenomenological Dyer-Roeder approach (or even some unknown procedure) should be preferred.

\subsection{Reconstruction of $\alpha$}

By extending the above discussion, it is natural to investigate the 
possibility to obtain $\alpha$ directly from the data, that is, 
without assumptions about its functional behavior. 
The extra bonus is that the difference between Dyer-Roeder and weak lensing predictions can be directly inferred. In the $\Lambda$CDM model, the Dyer-Roeder 
equation can be rewritten as: 

\begin{equation}
\alpha(z)\Omega_m= \frac{-E(z)}{1+z}\frac{d}{dz}\left[ 2(1+z)^2 
E(z) \frac{d}{dz} \frac{d_L}{(1+z)^2} \right] \frac{1}{d_L}. 
\end{equation}
Note also 
that the right side of this equation depends only on observational 
functions, and, as such, one may reconstruct the smoothness 
parameter for a general $\Lambda$CDM model, that is, regardless the 
values of the curvature parameter. 

\begin{figure*} 
\centerline{\epsfig{figure=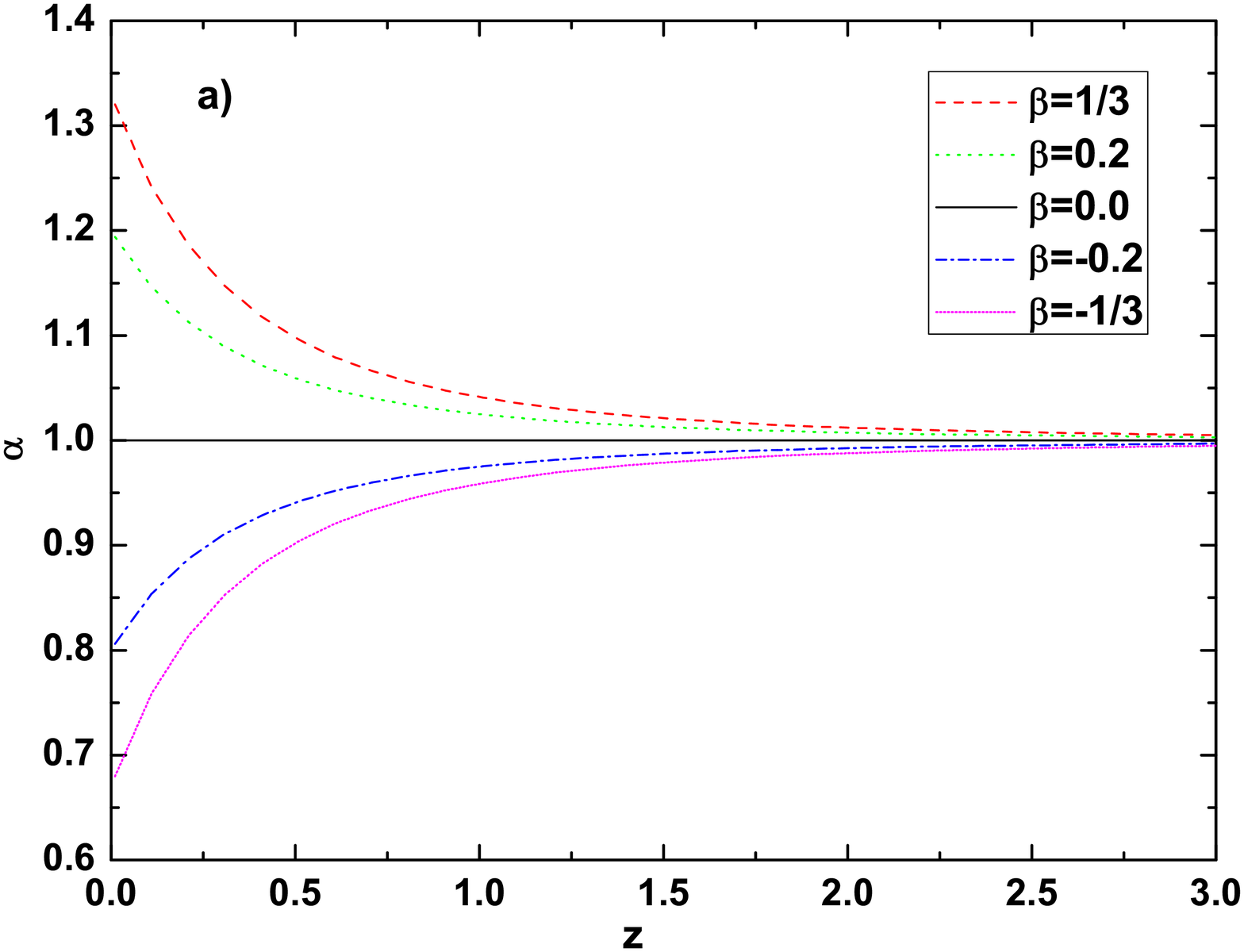,width=3.1truein,height=2.5truein} 
\epsfig{figure=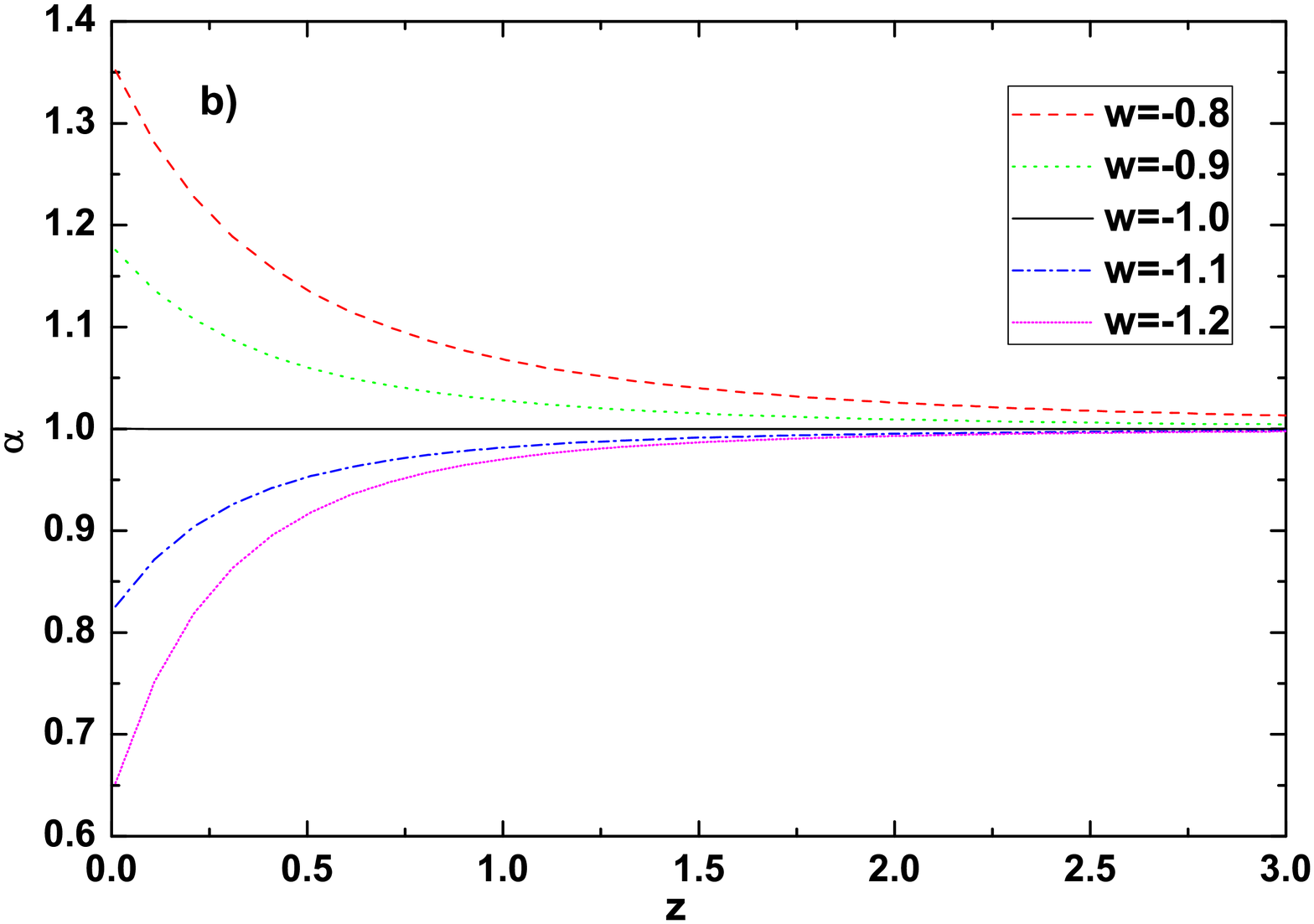,width=3.1truein,height=2.5truein} \hskip 
0.1in} \caption{$\Lambda$CDM versus XCDM cosmologies. 
Theoretical reconstruction of the smoothness parameter. In panel {\bf a)} we show the reconstruction considering 
a flat $\Lambda$CDM Universe with $\Omega_m=0.27$, $\gamma=1$ and several values for $\beta$. In panel {\bf b)}
a flat XCDM model was considered with $\Omega_m=0.27$, $\alpha=1$ and several values of $w$. Note that the same result can be  
obtained from different assumptions. Therefore,  in order to identify clearly the physical origin of the result the 
reconstruction should be performed in combination with other consistency checks.} 
\label{figrec} 
\end{figure*}

It is also worth notice that whether a smaller value of the left 
hand side is measured compared to independent estimates of 
$\Omega_m$, we have a constant $\alpha$. But what happens if a 
redshift dependence is detected? In principle, two effects may be 
present, the $\alpha$-effect or an unknown  deviation from 
$\Lambda$CDM. This is shown in Fig. \ref{figrec}, where in the 
left panel the $\alpha$-effect is displayed for the parametrization 
of the equation \ref{alphaz} and in the right panel a XCDM model is 
considered for $\alpha=1$ and different values for the dark energy 
equation of state $w$. We see that a XCDM model produces the same 
behavior as the parametrization considered for $\alpha$. 

Can we distinguish the possible effects? At present, the  best 
answer is that it depends on. For instance, if the ${\cal{C}}(z)$ 
test is zero and the reconstruction of $\alpha$ is not a constant, 
the true model cannot be $\Lambda$CDM. On the other hand, if the 
${\cal{C}}(z)$ is different from zero together with a departure from 
estimates of $\Omega_m$, probably, the $\alpha$-effect is playing 
the basic role. Of course, both effects may be at work, then 
different tests combined are required in order to identify what is 
truly happening at low redshifts. 

\subsection{Other Effects?}

So far we have analysed only the effects of the local 
inhomogeneities in the cosmological tests. But are there other 
effects taking place which were not accounted? One possibility is 
that the Etherington principle is not valid \citep{ETHER33}. This 
effect can change the distances as well, hence the cosmological 
tests will be affected. A full analysis of this effect will be 
published elsewhere.

\section{Conclusions}

In this Letter we have shown that cosmological tests originally 
proposed to find deviations from the FLRW metric or from a flat 
$\Lambda$CDM model are affected when distance measurements are used. 
This may happen due to the preferred lines of sight of the detected 
objects, e.g. SNe Ia, which results in a different distance from the 
standard FLRW approach. When this effect is taken into account, a 
new distance (sometimes called the Dyer-Roeder distance) is derived. In this 
approach, the effects of the local inhomogeneities are 
phenomenologically characterized by the smoothness parameter 
$\alpha$. It has been shown that if such a parameter is 
different from the unperturbed FLRW value ($\alpha=1$), artifacts 
are produced when the ``Copernican Principle'' 
is directly tested from observations.
 
It is also interesting that the fine tuned correspondence between the Dyer-Roeder and weak lensing approaches [suggested by \cite{bolejkodr2011}],  
implies that the consistency tests can also be used in reverse manner, that is, to probe the more realistic description of the small-scale inhomogeneities. 
This happens because the amplitudes of the deviations are heavily dependent on the adopted procedure (in certain sense,  the parametric Dyer-Roeder description encodes more possibilities).

We have also proposed  a method to reconstruct the smoothness 
parameter directly from the observations when a $\Lambda$CDM model 
is assumed. A discussion of how a different cosmology can affect the 
reconstruction was performed and it was recognized that different 
tests will be necessary in order to disentangle the cosmological 
model from the effects of the inhomogeneities (see Figs. 2a e 2b). Naturally, as happens with ${\cal{C}}(z)$ and ${\cal{L}}(z)$ tests, the reconstruction 
itself can also be used to identify the more realistic approach for describing the late time clumpiness effects. 

\acknowledgments

The authors are grateful to F. A. Oliveira, R. C. Santos, and F. Andrade-Santos for helpful discussions.  VCB is supported by CNPq and JASL is partially supported by 
CNPq and FAPESP (Thematic Project 04/13668-0).

\clearpage


\begin{thebibliography}{}

\bibitem[\protect\citeauthoryear{Alcaniz et al.}{2004}]{AL04} Alcaniz  J. S., Lima J. A. S., Silva R., 2004, Int. J. Modern Phys. D, 13, 1309 

\bibitem[\protect\citeauthoryear{Amanullah et al.}{2010}]{union2} Amanullah R. et al., 2010,  ApJ, 716, 712 

\bibitem[\protect\citeauthoryear{Bassett \& Kunz}{2004}]{Basset04} Bassett B. A., Kunz M., 2004, Phys. Rev. D, 69, 101305 

\bibitem[\protect\citeauthoryear{Bertotti}{1966}]{bert66} Bertotti B., 1966, Proc. R. Soc. London A, 294, 195 

\bibitem[\protect\citeauthoryear{Bolejko}{2011a}]{bolejko2011w} Bolejko K., 2011, A\&A, 525, 49 

\bibitem[\protect\citeauthoryear{Bolejko}{2011b}]{bolejkodr2011} Bolejko K., 2011, MNRAS, 412, 1937 

\bibitem[\protect\citeauthoryear{Busti \& Santos}{2011}]{hzdata} Busti  V. C., Santos R. C., 2011, Research in Astronomy and Astrophysics, 11, 637  

\bibitem[\protect\citeauthoryear{Busti et al.}{2012}]{bls2012} Busti  V. C., Santos R. C., Lima J. A. S., 2012, Phys. Rev. D 85, 103503, arXiv:1202.0449 [astro-ph] 

\bibitem[\protect\citeauthoryear{Clarkson et al.}{2008}]{clark08} Clarkson C., Bassett B., Lu T., 2008, Phys. Rev. Lett., 101, 011301 

\bibitem[\protect\citeauthoryear{Clarkson \& Zunckel}{2010}]{ex3} Clarkson C., Zunckel C., 2010, Phys. Rev. Lett., 104, 211301  

\bibitem[\protect\citeauthoryear{Clarkson et al.}{2011}]{clark11} Clarkson C., Ellis G., Faltenbacher A., Maartens R., Umeh O., Uzan J.-P., 2011, arXiv:1109.2484[astro-ph]

\bibitem[\protect\citeauthoryear{Dyer \& Roeder}{1972}]{Dy72} Dyer C. C., Roeder R. C., 1972, ApJ, 174, L115 

\bibitem[\protect\citeauthoryear{Dyer \& Roeder}{1973}]{Dy73} Dyer C. C., Roeder R. C., 1973, ApJ, 180, L31 

\bibitem[\protect\citeauthoryear{Ellis}{2009}]{rev_ellis} Ellis G. F. R., 2009, J. Phys. Conf. Ser., 189, 012011 

\bibitem[\protect\citeauthoryear{Etherington}{1933}]{ETHER33} Etherington I. M. H., 1933, Phil. Mag., 15, 761 

\bibitem[\protect\citeauthoryear{February et al.}{2010}]{render} February S., Larena J., Smith M., Clarkson C., 2010, MNRAS, 405, 2231 

\bibitem[\protect\citeauthoryear{Gunn}{1967}]{gunn67} Gunn J. E., 1967, ApJ, 150, 737 

\bibitem[\protect\citeauthoryear{Holanda et al.}{2010}]{Holanda10} Holanda R. F. L., Lima J. A. S., Ribeiro M. B., 2010, ApJ, 722, L233, arXiv:1005.4458 [astro-ph.CO]  

\bibitem[\protect\citeauthoryear{Holanda et al.}{2011}]{Holanda11} Holanda R. F. L., Lima J. A. S., Ribeiro M. B., 2011, A\&A, 528, L14, arXiv:1104.3753 [astro-ph.CO]  

\bibitem[\protect\citeauthoryear{Jordan et al.}{1961}]{Sachs2} Jordan P., Ehlers J., Sachs R. K., 1961, Akad. Wiss. Mainz, 1, 1 

\bibitem[\protect\citeauthoryear{Kainulainen \& Marra}{2009}]{marra2009} Kainulainen K., Marra V., 2009, Phys. Rev. D, 80, 123020

\bibitem[\protect\citeauthoryear{Kantowski}{1969}]{Kant69} Kantowski R., 1969, ApJ, 155, 89 

\bibitem[\protect\citeauthoryear{Linder}{1988}]{linder88} Linder E. V., 1988, A\&A, 206, 190  

\bibitem[\protect\citeauthoryear{Linder}{1998}]{linder98} Linder E. V., 1998, ApJ, 497, 28   

\bibitem[\protect\citeauthoryear{Linder}{2003}]{Lin03} Linder E. V., 2003, Phys. Rev. Lett., 90, 091301 

\bibitem[\protect\citeauthoryear{Mattsson}{2010}]{mattsson2010} Mattsson T., 2010, Gen. Relativ. Gravitation, 42, 567 

\bibitem[\protect\citeauthoryear{Padmanabhan and Choudhury}{2003}]{Pad03} Padmanabhan T., Choudhury T. R., 2003, MNRAS, 344, 823 

\bibitem[\protect\citeauthoryear{Sachs}{1961}]{Sachs61} Sachs P. R. K., 1961, Proc. R. Soc. London A, 264, 309  

\bibitem[\protect\citeauthoryear{Sahni \& Starobinsky}{2006}]{ex2} Sahni V., Starobinsky A. A., Int. J. Modern Phys. D, 15, 2105  

\bibitem[\protect\citeauthoryear{Sahni et al.}{2008}]{sss2008} Sahni V., Shafieloo A., Starobinsky A. A., 2008, Phys. Rev. D, 78, 103502 

\bibitem[\protect\citeauthoryear{Saini et al.}{2000}]{ex1} Saini T. D., Raychaudhury S., Sahni V., Starobinsky A. A., 2000, Phys. Rev. Lett., 85, 1162  

\bibitem[\protect\citeauthoryear{Santos \& Lima}{2008}]{SL07} Santos R. C., Lima J. A. S., 2008, Phys. Rev. D, 77, 083505, arXiv:0803.1865 [astro-ph] 

\bibitem[\protect\citeauthoryear{Santos et al.}{2008}]{SCL08} Santos R. C., Cunha J. V., Lima J. A. S., 2008, Phys. Rev. D, 77, 023519, arXiv:0709.3679 [astro-ph]

\bibitem[\protect\citeauthoryear{Uzan et al.}{2008}]{ellis2008} Uzan J.-P., Clarkson C., Ellis G. F. R., 2008, Phys. Rev. Lett., 100, 191303 

\bibitem[\protect\citeauthoryear{Wei}{2010}]{hymnium} Wei H., 2010, J. Cosmology Astropart. Phys., 1008, 020 

\bibitem[\protect\citeauthoryear{Zel'dovich}{1964}]{Ze64} Zel'dovich Ya. B., 1964, Sov. Astron., 8, 13  

\bibitem[\protect\citeauthoryear{Zunckel \& Clarkson}{2008}]{zc2008} Zunckel C., Clarkson C., 2008, Phys. Rev. Lett., 101, 181301  






\end{thebibliography}
\end{document}